# Challenges with Automation in Digital Forensic Investigations

Joshua I. James and Pavel Gladyshev
{joshua.james, pavel.gladyshev}@UCD.ie

Digital Forensic Investigation Research Group
University College Dublin
Belfield, Dublin 4
Ireland

## Abstract

The use of automation in digital forensic investigations is not only a technological issue, but also has political and social implications. This work discusses some challenges with the implementation and acceptance of automation in digital forensic investigation, and possible implications for current digital forensic investigators. Current attitudes towards the use of automation in digital forensic investigations are examined, as well as the issue of digital investigators' knowledge acquisition and retention. The argument is made for a well-planned, careful use of automation going forward that allows for a more efficient and effective use of automation in digital forensic investigations while at the same time attempting to improve the overall quality of expert investigators. Targeting and carefully controlling automated solutions for beginning-investigators may improve the speed and quality of investigations while at the same time letting expert digital investigators spend more time utilizing expert-level knowledge required in manual phases of investigations. By considering how automated solutions are being implemented into digital investigations, investigation unit managers can increase the efficiency of their unit while at the same time maximizing their return on investment for expert-level digital investigator training.

**Keywords:** Digital Forensics, Investigation Automation, Push Button Forensics, Training, Education, International Collaboration

## 1. Introduction

Highly automated digital forensics – sometimes referred to as "push-button forensics" (PBF) – receives much criticism from the digital investigation community. Criticisms generally appear to focus on two aspects of digital investigations: a deterioration of expert knowledge by an over reliance on PBF, and a perceived less thorough, or lower quality, investigation when relying on a high level of automation. However, critics also currently accept a certain level of automation to help them in their daily tasks. Manually comparing each hash in a hash database, for example, would be impractical without some level of automation. The challenge comes when higher-level processes, such as analysis, are being automated, and also when the investigator begins to loose understanding of the underling concepts of the investigation. Both of these scenarios are currently happening to some extent. Digital investigation software suits such as EnCase ([Guidance 2010](#)), Forensic Tool Kit ([AccessData 2010](#)), Autopsy Forensic Browser ([Carrier 2010](#)), and others allow an investigator to conduct preliminary, and even some complex investigation tasks simply by knowing which button to press. These popular tools endeavor to make the job of the investigator easier, or even remove the expert altogether, as seen in a claim from Access Data ([2009](#)): "Digital investigations are no longer the exclusive domain of highly trained experts". Full-featured forensic software suites are not the only potential source of issues. Simple programs such as TraceHunter ([Zhu, James et al. 2010](#)) provide correlation, interpretation, and some analysis of the Windows Registry regardless of the practitioner's knowledge, and the same can be said for scripts written by experienced investigators that are then distributed to others who may have little to no knowledge of the underlying processes and data sources. Despite criticisms, the reality today is that investigators are currently using a high level of automation in investigations, and newer, or uninterested, investigators who are trained on a specific tool may be unable or unsure how to do investigations without the use of automation. This is contrary to the view that every investigator should have a solid, non-tool-centric knowledge of the investigation process. Whether this lack of knowledge comes from a lack of training, time or funding, it has real implications on the quality of investigations being conducted.

## 2. Related Work

Kovar ([2009a](#); [2009b](#)), considers the value of push button forensics, and cites three main reasons for the acceptance of increased automation: The tool vendors' need for the expanded non-expert market will evolve push-button interfaces, speed-related financial interest from consumers of computer forensics services (law enforcement, private sector, etc.), and the growing volume of digital evidence resulting in case backlogs. The general reply to this claim was that practitioners must understand "… the science, logic, and art behind the PBF tools" ([Kovar 2009b](#)). The admissibility of evidence that was the result of a lesser-trained technician running a tool was also questioned. Kovar claims that many technicians are already using push button tools as part of the computer forensic process. He ultimately concludes, "There clearly are risks to using PBF and inexperienced examiners inappropriately, but through sound business practices they can safely contribute to our projects and improve our efficiency in the process". However, no mention is given to how PBF tools should be implemented in the investigation process to ensure quality. Kovar is also focusing only on inexperienced examiners, whereas improperly implemented automation can potentially become an issue for experienced and inexperienced investigators alike, if not managed properly.

      Similarly, Casey ([2006](#)) states that "too little knowledge is a dangerous thing"

in regards to digital investigations. He claims that inexperienced internal or outsourced investigators may have an "…over reliance on user-friendly or automated forensic software", and may "…apply a form of pseudo-automation by rigidly following predefined protocols". Casey also states that, "Inexperienced individuals who do not critically review the results of a tool will inevitably misinterpret or completely miss digital evidence". He cites the limitations of self-assessment, and calls for set standards in digital investigation competencies and processes.

While each point may be valid, the reason for the use of less-experienced investigators, reliance on automation, and a lack of certification and thorough checks of labs and contractors is simply an issue of time and cost. Some organizations may have the resources to be able to use certified labs and highly trained and experienced investigators, but law enforcement is a trade off between the cost and results, known as a 'balance scorecard' (Ashworth 2010). "[Hiring specialists] would force law enforcement agencies to incur great expense, perhaps a crushing expense for smaller police departments that already face tremendous budget pressures (US v. Comprehensive Drug Testing, 2009)." If Law Enforcement (LE) gets "satisfactory", admissible results from a non-certified lab or automated software in half the time and cost, then management will have to judge whether the cost savings outweigh the perceived cost to justice, which also raises the challenge of precision measurement in digital investigations. This is an unfortunate reality that may be helped with improved performance measurements and standardization, such as the standards proposed by Lee (2008) and the UK Forensic Science Regulator (2011), but will never be completely solved; potentially resulting in missed digital evidence.

Another challenge is the definition of an experienced investigator. Experience is not the same as competence. Irons, Stephens, et al. (2009) specifically differentiate between practice and theory as well as skills and knowledge when dealing with the training of competent digital investigators, claiming that each area needs to be developed to ensure competency. However, neither Irons nor the UK Forensic Science Regulator focus on the investigator's retention of learned theory and knowledge; only present performance and undefined technical skill maintenance. Similarly, most studies assume that all investigators or technicians are also interested in their job, and are free of possibly undiagnosed psychiatric and learning disorders that can inhibit or prevent learning and retention beyond the basics required for the job[1] (Goldstein 1997; Wender, Wolf et al. 2001). This is a bold assumption considering "only 51 percent [of Americans] now find their jobs interesting" (CBS 2010), and approximately 1 in 4 adults in America suffer from a mental disorder, some of which may affect learning and retention (Kessler, Chiu et al. 2005). In these situations, potential evidence may be missed due to disinterest or lack of retention of knowledge.

Despite skepticism, automation is already being used in digital investigations. Digital forensic triage, for example, is a rapidly growing and highly automated area. Many works (Rogers, Goldman et al. 2006; Casey, Ferraro et al. 2009; Goss and

---

[1] For more information on workplace disability anti-discrimination legislation in the US and UK, see the Americans with Disabilities Act of 1990 (ADA) and the Disability Discrimination Act 1995 (DDA)

Gladyshev 2010; Koopmans 2010; Mislan, Casey et al. 2010) have called for advances in computer and mobile phone triage because they realize the benefits of fast, automated, on-scene intelligence gathering. These benefits have been examined within a UK high-tech crime unit in Goss and Gladyhsev (2010), which showed a reduction in the amount of seized computers and suspect data needing an in-depth analysis. Goss also compared automated triage performance with manual investigation finding that triage gave comparable examination results in a fraction of the time for specific case types where in-depth knowledge is not required, such as child exploitation image detection. In more complicated cases, where knowledge of the system is necessary for analysis, triage was less reliable, and often missed potential evidence.

While triage has been shown to have definite benefits in some specific cases, there are still challenges, such as investigator training, potential missed evidence and verification challenges that need to be addressed.

## 3. Knowledge

Ianuzzi (2007) claims that one challenge with automating digital investigations is that automation may "dumb down the profession". This parallels the idealistic view, shared by some investigators and researchers, that most investigators have already achieved an undefined "good" standard in forensic knowledge. This seems to be a dangerous assumption, but is commonly cited as an argument against automation. There are currently unqualified, unknowledgeable forensic practitioners that are conducting digital investigations (Jones 2004; Everett 2005). This may be true for practitioners who were chosen for the job because they had basic computer skills, to people with 10+ years experience that generally have no interest in learning anything but the minimum required to keep their job.

Just like any profession, expansive expert knowledge cannot come only from the field; it must also come from continued formal and informal study. As stated by Gogolin (2010), "digital skills are perishable if not kept current", and 75% of investigators receive between zero (30% of the sample) and 5 days annual training. Casey (2009) affirms that "there are some fundamental principles, concepts, and skills that everyone in this field must abide by and know". Likewise, Irons, Stephens et al. (2009) claim, "the implicit expectation is that digital investigators should be competent before undertaking any digital investigation duties". But this is not always the reality. Many investigators have little to no digital investigation training before starting, and even after, "[o]nly 34% of [digital forensic] investigators [surveyed in Michigan, USA] received formal training in laboratory forensics, with the majority being trained 2 weeks or less" (Gogolin 2010).

Likewise, EURIM (2004) claimed that in 2004 "…barely 1,000 [police officers in the UK] have been trained to handle digital evidence at the basic level and fewer than 250 are currently with Computer Crime Units or have higher level forensic skills". Neil Hare-Brown claimed that as much as 20% of digital investigation consultants in the UK are "unfit" (Everett 2005); however, "fitness" in this case is undefined, and many initiatives since 2005 may have improved this estimation. This lack of definition for investigator fitness is industry-wide since "there is an absence of standards and competencies in the field of cybercrime" (Jones 2004). However, efforts for standardization are being made (Lee 2008; HomeOffice 2011). Also, consider that these statistics are from the US and Europe; areas that generally have funds to invest in standards and training. This may mean that countries with less funding receive even less quality training, and may have less access to education, equipment and skilled investigators in general. The result of this global, and even national, imbalance in funding, training, aptitude, interest, intelligence and skill results in few experts, many technicians and a huge variation in the quality of digital investigations world wide.

A lack of knowledge can lead to missed evidence and incorrect conclusions, and based on the given statistics this is likely happening without the use of automation. A "dumbing down" of the profession may happen if investigators with expert level knowledge over-rely on automation for their analysis, but for untrained investigators with little-to-no knowledge to begin with, automation may enable them to find and evaluate evidence that may have otherwise been overlooked.

## 3.1 Digital Investigation Training

In a report by the High Technology Crime Investigation Association (HTCIA) (2010), training for practitioners was a primary concern, claiming that 73% of respondents believe they do not receive enough training, especially in digital forensics, online investigations, and computer and network security. "Many respondents did not indicate that more personnel were necessarily needed, but rather believed more training at all levels was important." This parallels a study done by Gogolin (2010) that showed the majority of digital investigators interviewed were not receiving formal training before or during their employment. The result is that "…it typically takes between one and two years of on-the-job training before a newly minted forensics examiner is proficient enough to lead an investigation" (Garfinkel 2010). These studies indicate that more training is desired, but there are some challenges, especially funding, time and training quality, that are hindrances to investigator development.

### 3.1.1 Training Quality and Retention

One challenge is simply a lack of quality investigator training, partially caused by the absence of a "…common agreement on the sets of skills for which training would be most beneficial" (EURIM-ippr 2010). Without a standard digital investigation common body of knowledge (CBK), such as the work proposed by Lee (2008), colleges and other training organizations are able to design digital investigator training based on marketing rather than industry-driven quality. Many organizations have started programs specifically to capitalize on the current popularity of crime investigation TV shows, and the increased interest in the field based on the show, known as the "CSI effect". "The CSI effect is leading many high school and college students to take forensic science or criminalistics courses and prepare for careers in the field (Dempsey and Forst 2009).” When education programs have no standard to which they must adhere, it becomes difficult for new students, and possibly employers, to ensure they are getting a quality education. A standard body of knowledge or standard of quality, however, should not be confused with an educational content limitation. The aim of education is to prepare independent thinkers, but those thinkers should also be well versed in their respective areas. This is not always the case. As stated by Jones (2004), "we all know of training courses where there is no independent assessment of the student knowledge, yet the certificates issued to the students (after payment of the course fee) lend them credibility".

The same can also be said for vendor training with no third-party assessment where the certification of an unfit customer and what is best for the vendor's business could be a conflict of interest. Also, in other fields it has been shown that short, intensive training sessions, such as those offered by many certification bodies, produce gains in knowledge that are acceptable to immediately certify, usually at the end of the training session (McGguire, Hurley et al. 1964; Wik, Myklebust et al. 2002), but the majority of the knowledge is not retained after 6 months, as illustrated in Figure 1. "The data strongly suggest that in the absence of opportunity to practice the recently enhanced skill under conditions where critical evaluation of performance is available, individual gains are not maintained over a period of several months (McGguire, Hurley et al. 1964)." Since some basic and many advanced forensic concepts are not used on a daily or even weekly basis, intensively trained and certified investigators could eventually have credentials that certify them beyond the scope of their understanding or skill. This situation is made worse when a trained investigator is able to rely on highly automated tools, rather than utilizing their own knowledge. Automation in this case may allow unknowledgeable investigators to appear acceptable, and may cause a deterioration of knowledge in knowledgeable experts.

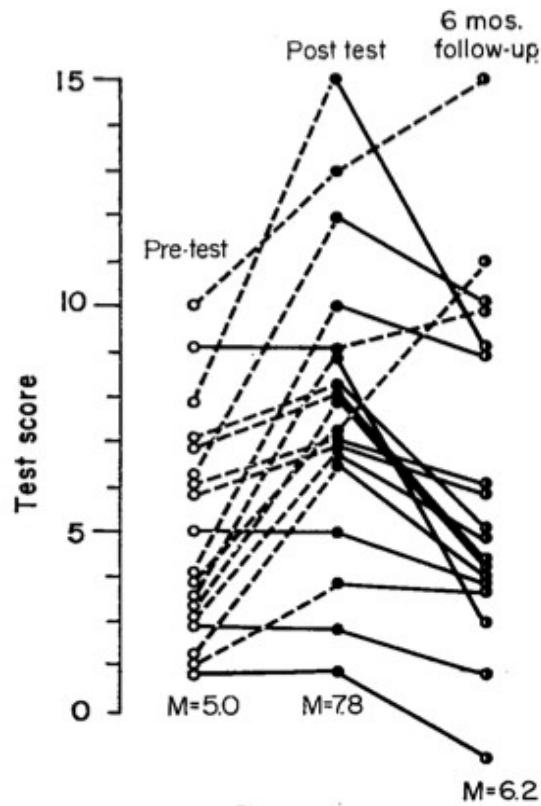

*Figure 1. A diagram of scores before, immediately after and 6 months after a short period intensive learning, showing immediate gains and long-term retention ([McGguire, Hurley et al. 1964](#)).*

*3.1.2 Time*

Along the lines of intensive training courses, time for training is a factor. Consider that in 2004 the United Kingdom digital crime investigation backlog was 6 to 12 months ([EURIM-ippr 2004](#)), and rose to 18 to 24 months in 2009 ([2009](#)) before being improved through a number of policy, case prioritization and evidence outsourcing initiatives ([Kohtz 2011](#)). Currently in the United States there are reports of backlogs from 12 to 18 months ([Raasch 2010](#)), and in some cases "approaching or exceeding 2 years" ([Gogolin 2010](#)). With a growing backlog, investigators are pressured to work faster, which could mean less time to study underlying concepts, even informally. Likewise, during discussions few investigators have reported being up to date on research topics that could improve their processes. If a technician can quickly be trained to extract information without receiving in-depth training about the underlying concepts, and the information they extract is still admitted into court, then, as previously stated, management will have to judge whether the time and cost savings outweigh the perceived cost to justice.

*3.1.3 Funding*

Funding is yet another rival to knowledge. "Most cybercrime training funding comes out of budgets rather than assistance [grants]" ([HTCIA 2010](#)), which means needed training is competing with department maintenance, new equipment, software, budget cuts, increasing amounts of data as well as expanding job scope. "No agency in the study reported an increase in funding for digital crime investigation over the previous year ([Gogolin 2010](#))." Casey, Ferraro et al. ([2009](#)) observes "few [Digital Forensic Laboratories] can still afford to create a forensic duplicate of every piece of media and perform an in-depth forensic examination of all data on those media". This may force departments to choose purchasing a software suite and rely on its automated features, rather than have trained experts but not enough software licenses.

A common training model appears to be purchasing a certain tool, sending one person to receive training, and then having this person train others when they return. This method could be effective, but as seen before, a lack of understanding or knowledge retention could lead to partial or incorrect knowledge being disseminated. Again, this is considering countries that have a budget to work with. In some countries the total annual budget for cybercrime investigation may be less than the cost of a single three-day training course from a popular forensic software vendors.

Ultimately, in the current investigation model, tools with more features and a greater amount of automation will save time, but will consistently need to be upgraded and renewed. More tools will be needed to cover the expanding scope, and new equipment will be needed to support the tools. The technological needs of a department contend with the training needs, and products that require little to no training to use will be considered a cost saver.

### *3.1.4 Interest and Attitudes*

Interest in the job and belief in its mission, as mentioned before, is also often taken for granted. This has a number of implications: First, if people put their personal desires before the good of the company, funding can be easily wasted, as seen in the US Congress ([2004](#)) where the cost of excessive flights outweighed the benefit of attending a conference. Second, with a mere 51% job satisfaction rate in the US ([CBS 2010](#)), and a +34% net job satisfaction rate in the UK ([CIPD 2010](#)), it would be naïve to think that law enforcement, and digital investigators specifically, are immune to uninterested employees. This means some investigators may be doing the minimum required for the job, not for justice, but simply as a means to an end. Finally, overexposure to particular types of investigations could lead to categorization bias ([Anderson 1991](#)). Each of these implications could have a profound negative impact on the quality of investigations being conducted. By utilizing automation in digital investigations as it is currently implemented, uninterested persons may be able to conduct investigations with minimal effort that appear satisfactory but in fact lack in-depth expert analysis.

### *3.2. Certification and Licensing*

Certification and licensing is a much-debated topic in information security and digital investigation. In the US, for example, some states require digital forensic examiners to be licensed as private investigators, while others do not. And "licensing requirements for forensic examiners have yet to be standardized on a national level…" ([Manes and Downing 2009](#)), which can lead to examiner licensing issues in multi-state cases. The American Bar Association ([2008](#)) is opposed to requiring private investigator licenses for forensic examiners, but "supports efforts to establish professional certification or competency requirements for such activities…". Other countries have yet to adopt licensing requirements for digital forensic examiners, and currently rely on the plethora of digital examiner certification programs. But again, training and certification will only be done in countries with the budgets to do so. And in those countries, without a defined standard, investigators and employers may find it difficult to differentiate between the levels of quality of various certifications.

*3.2.1 Investigator Certification*

The Digital Forensics Certification Board (DFCB), SANS, (ISC)², International Association of Computer Investigative Specialists (IACIS), The International Society of Forensic Computer Examiners (ISFCE), EC-Council, forensic software vendors, and many others offer certification for digital investigators. Some with little or no work experience requirements. In *Codes of Practice and Conduct for forensic science providers and practitioners in the Criminal Justice System* (2011) the term "competence" is often used, but nowhere is competence or competence testing explicitly defined, save references to National Occupational Standards (NOS) from Skills for Justice (2010), which claim to "describe competent performance in terms of outcomes [and] …allow a clear assessment of competence against nationally agreed standards of performance". However, NOS appear to be limited to overall general forensic procedures, allowing for a standard procedure to ensure the proper consideration and handling of evidence, but with no digital examination-specific performance criteria. Beyond NOS, the Forensic Science Regulator (2011) has proposed to rely on "appropriate records of education, training, skills and experience in sufficient detail to provide evidence of proper training and formal assessment", which again leaves the definition of "appropriate" to each individual department. This lack of concrete definition, combined with an overwhelming amount of available certifications, can cause confusion among practitioners and employers, and allow incorrect evaluation of competency. "It's important to understand that certification does not mean mastery… In fact, certification doesn't necessarily even mean professional *competency* (Huber 2010)."

*3.3. International Collaboration*

Forensic practitioners and government officials often talk about LE knowledge, standards and competency from a national or regional perspective. However, Huber (2010) explained that "digital forensics is very much an interstate and international issue". The IT security sector is beginning to accept the need for global cooperation for incident response (Sambandaraksa 2010), but the forensics community has been slower in this regard. Some attempts at international cooperation involving investigator-sharing for digital forensic investigations have taken place (INTERPOL 2008), which have led to many questions of jurisdiction and acceptable procedure. However, investigations with cross-jurisdiction components happen often, with varying degrees of success. According to Brenner (2006) "countries' ability to assert jurisdiction over those who perpetrate cybercrime, both locally and transnationally, does not seem to be particularly problematic. …[The] issue is how to prioritize conflicting claims to assert jurisdiction over transnational cybercriminals". When considering standardization, unless there is an accepted global standard for digital investigations, the way evidence is handled in another country may not meet the legal or time requirements of requesting countries. Even if standards were put in place, practitioners in some countries may not always be able to meet these standards because of a lack of training, funding, etc. For this reason international collaboration is imperative to create non-conflicting standards. Global participation and investment will help build legal and technical infrastructures worldwide, and unified standards encapsulated in formally proven automated tools will help to streamline international investigations, for trained and untrained alike, which may ultimately have a positive impact locally.

## 4. More Intelligent Automation

Since the definition of Digital Forensic Science at the Digital Forensics Research Workshop in 2001, the field has grown almost as dramatically as technology itself. As described by Casey (2009), digital forensic science is "coming of age", which not only brings about a maturation in the principal concepts of the field, but also an increased scrutiny against these principles and their lack of rigorous scientific backing. Investigators are now finding themselves overwhelmed with the scale and quantity of cases, along with the pressure of increasingly restrictive standards. This combination simply translates into a consistently increasing number of delayed or even neglected cases. To combat an "impending crisis in digital forensics", Garfinkel (2010) claims "advanced systems should be able to reason with and about forensic information in much the same way that analysts do today". This means automated processes that can interpret what observed information means rather than simply interpreting bits to human-readable information, as is currently being done.

*4.1 Current State of Automation in Digital Forensic Investigations*

As mentioned before, automation is already being used in digital investigations, and research is being done in this area. A high level of automation can already be achieved with evidence collection, processing, and to some extent, documentation, and a growing amount research is being done that attempts to automate analysis (Khan, Chatwin et al. 2007; Zhu, James et al. 2009; James, Gladyshev et al. 2010). This work tends to move from simply representing data as human-readable information to representing what the information means, and even reasoning about this meaning. Despite this, current tools still focus on converting data to information that an investigator can then use to manually draw conclusions. No software is perfect, and the ability for non-professional programmers to add on to this software allows even more room for error. Further, James and Gladyshev (2010) found that only 23.3% of respondents verified the accuracy of examinations. From this as well as other discussed topics, it is likely that verification of the software is not always properly done, and sometimes not at all. Furthermore, verification of the non-professional add-ons could be regularly overlooked.

Regardless, no mention is usually given to what evidence is currently being missed during investigations. Because of budget restrictions, some departments only have and know how to use one piece of "full-function" software. This attitude shows an over-reliance on a specific piece of software. And this is where the one challenge for justice comes in; when investigators are previewing a system, they must get to a point where they decide if the case needs further analysis or not. Since time, money and justice are always competing, many technicians are taught how to run automated tools and then analyze the results themselves. If it can be assumed that the automation built into these tools is reliable – keyword list search, hash search, etc. – then there are really two challenges that remain: a reliance on the investigator to correctly run the automated tool and a reliance on an investigator, with possibly no knowledge of the data, to interpret the presented information.

Currently there are investigators that are not running all of the automated tools available to them. For example, during discussions, some examiners who make the decision to continue or stop the examination claimed they never use known hash databases in exploitation cases because hash comparisons "never find anything", and is therefore not worth the additional time. Both of these are challenges of knowledge, and perhaps policy, which have been discussed previously, but it is also about investigator attitudes. These challenges appear to be partially caused by the over-use of automation itself, but the issue is really what is automated, where automation is placed, and who is using it.

## 4.2 Opportunities with Automation in Digital Forensic Investigation

There are a number of opportunities with a higher level of automation at the preliminary analysis level. These include:

- **Standardized knowledge and investigations**

    Automation can be used for arduous manual tasks, such as hash matching, but also to encode the knowledge of trained investigators in a repeatable, verifiable way. "[C]omputer forensic software should ideally provide an objective and automated search and data restoration process that facilitate consistency and accuracy ([Guidance 2009](#))." If automated first-responder processes were based on expert knowledge and standardized processes, higher quality first-level investigations could take place than normally would with a less trained technician.

    Automation is also an accumulation of expert knowledge. This bank of knowledge applied at the first step in the investigation may lead to less missed evidence by encoding knowledge of obscure evidence that a normal investigator would not necessarily know about. Not only does automation allow for standardized processes in a department, but the same standard of first-level investigation could take place everywhere in the world regardless of investigator or department knowledge and budgets. This could allow for first-level digital investigations in countries that would otherwise not have the capability. Ultimately, an automated first-level analysis applied at the first-responder stage

of the investigation may allow for a faster, higher quality, standardized preview with little associated training investment.
- **Department load reduction and knowledge retention**
  Tasking the first responder with the automated first-level analysis benefits expert forensic investigators in a number of ways. First, since the preliminary analysis is being done in the field, less suspect machines will need to be analyzed by experts, as shown with current preliminary analysis techniques in (Goss and Gladyshev 2010). This reduction in suspect machines will reduce the workload on the forensic lab. Along the same lines, every suspect machine given to the expert will definitely require a thorough analysis. This means that valuable expert investigators will no longer need to simply run automated tools, but will normally be conducting investigations that will require in-depth knowledge. "Field triage, evidence previews and even rudimentary evidence collection can free investigators and forensic examiners to focus on investigative and analysis activities (HTCIA 2010)." Not only could there be a reduction of suspect devices, but potentially a reduction in cases that require a laboratory examination at all. Many digital forensic investigators as well as the Mac Forensics Lab (2010) agree that since the first responder is obtaining information and potential evidence, it could allow the first responder to immediately secure a confession based on results found on-scene. This is, however, only if the officer understands that what he or she is looking at is, in fact, relevant to the case. For this, there needs to be an understanding about how crime is committed.
- **Increased Training and Education**
  Further, with the reduction in workload and the need for in-depth expert knowledge for all incoming suspect material, investigators will, potentially, have more time and much more need for quality digital forensic investigation education. Since the experts will be using high-level expert knowledge more than before, more information will be retained providing a higher return on investment and more knowledgeable, competent experts.

## *4.3 Challenges with Automation in Digital Forensic Investigation*

The main challenge to automated tools is that they have neither a complete knowledge nor capacity to process this knowledge in a particular case. Because of this, if they do not show anything it does not mean that the evidence is not there. Ianuzzi (2007) and Goss and Gladyshev (2010) believe that completely automated process cannot work in all cases; automated tools will miss some evidence. Missed evidence, however, is not only a problem of automation, but instead a problem of investigation in general. Investigators also cannot know how much evidence they are missing.

   To approach this challenge, it must be known how well automated tools currently work, and in what situations they are best suited. Once a performance measure has been established based on the objectives of an investigation, weaknesses in automation may be objectively evaluated and improved upon. This must be able to include factors such as weight of the evidence in relation to the case, and accuracy of interpretation.

The second challenge is of the admissibility of automatically derived evidence by a non-expert. The UK Forensic Regulator (2011) has outlined that forensic software must be validated using specified, rigorous validation methods. Validation generally establishes that "the validation work is adequate and has fully demonstrated compliance of the method with the acceptance criteria for the agreed specification; and the method is fit for its intended use". Once an expert validates the tool, any user should then follow the tool's defined operating procedure. A non-expert investigator that followed the procedure, however, may be called to give expert testimonial about their findings. If they cannot establish that they have complete knowledge of how they arrived at their conclusions, this may introduce doubt and reduce the credibility of evidence derived using automated processes.

## 5. Implementing Advanced Automation in Digital Investigations

Automation may "dumb down" the profession when experts are able to rely solely on automated tools for all data interpretation and analysis tasks, and derived evidence is admitted in court without challenge. However, the technologies to be able to fully automate complicated evidential analysis and reasoning tasks without human intervention are not yet available, and courts admitting automatically extracted evidence without some sort of human expert verification appears unlikely.

Automation can potentially increase the speed of the investigation process, and reduce the number of suspect devices in the lab, which will ultimately reduce case backlog while avoiding bias and prejudice. Because speed is one of the biggest concerns during an examination, digital investigators need to re-evaluate the examination process, and consider faster ways to determine if an in-depth analysis is necessary, such as profiling (Marrington, Mohay et al. 2010) or automatic event reconstruction (Gladyshev 2005; James, Gladyshev et al. 2010).

Along with speed, the validity of automated processes must be formally proven. "Despite having a variety of practical techniques and tools, [Digital Forensics] provides little theoretical basis to support correctness of investigation findings (Gladyshev 2004)." Until the forensic community embraces formal methods, time consuming ad-hoc verification will continue, and the quality of the tools and investigations will suffer. Formal verification will also allow for the use and design of more efficient and intelligent forensic tools as called for by Garfinkel (2010). Finally, improved measurements in regards to the identification of digital evidence are necessary. Until baseline performance measurements are created for digital investigations that account for evidence identification in the context of the case, performance of highly automated tools will be unverifiable, and perhaps hinder development.

# 6. Conclusions

The argument that automation will "dumb down" the field of digital forensic investigations is only the case when automation is implemented poorly at the wrong phase of the investigation. When a trained, expert forensic examiner can do their job with the click of a button, that is the when knowledge and quality will be lost. Automation implemented correctly, at the correct phase of the investigation can produce equal or better results compared to a human investigator by conducting expert-level preview in a standardized way with less training for previewers. What is needed is a well-planned, careful use of automation going forward that allows for a more efficient and effective use of automation in digital forensic investigations while at the same time attempting to improve the overall quality of expert investigators.

Automation can increase the speed of the investigation process, and reduce the number of suspect devices in the lab, which will ultimately reduce case backlog while avoiding bias and prejudice. However, several issues still exist that need to be addressed before automation can increase the speed, reliability and effectiveness of digital investigations. These include improving the measurement of current investigation accuracy, pushing for formal verification, and reevaluation of what evidence is necessary to make informed decisions. Once done, push-button forensics could allow for faster standardized investigations, not only in countries that can afford expensive forensic software and training, but also all over the world.